\documentclass[12pt]{iopart}
\usepackage{cite}
\usepackage[english]{babel}

\begin{document}
\title{Effect of dipolar interactions on optical nonlinearity of two-dimensional nanocomposites}
\author{Andrey V. Panov$^{1,2}$}
\ead{panov@iacp.dvo.ru}
\address{%
$^1$Institute of Automation and Control Processes,
Far East Branch of Russian Academy of Sciences,
5, Radio st., Vladivostok, 690041, Russia\\
$^2$School of Natural Sciences, Far Eastern Federal University, 8, Sukhanova st., Vladivostok, 690950, Russia}


\begin{abstract}
In this work, we calculate the contribution of dipole-dipole interactions to the optical nonlinearity of the two-dimensional random ensemble of nanoparticles that possess a set of exciton levels, for example, quantum dots. The analytical expressions for the  contributions in the cases of TM and TE-polarized light waves propagating along the plane are obtained. It is shown that the optical nonlinearity, caused by the dipole-dipole interactions in the planar ensemble of the nanoparticles, is several times smaller than the similar nonlinearity of the bulk nanocomposite. This type of optical nonlinearity is expected to be observed at timescales much larger than the quantum dot exciton rise time. The proposed method may be applied to various types of the nanocomposite shapes.

The final publication (Andrey V Panov, 2013, J. Opt. 15, 055201) is available at http://iopscience.iop.org/2040-8986/15/5/055201, \\DOI: 10.1088/2040-8978/15/5/055201

\end{abstract}

%
\maketitle

In recent years, much attention has been given to optical nanostructures
containing, for instance, semiconductor quantum dots. Frequently, such
nanocomposites are organized as two-dimensional structures
\cite{Zimnitsky07,Bai09,Xiong10,Xuan11}. Normally, these structures are
dense-packed systems of nanoparticles. The high concentration of
nanoparticles will require taking into account the contribution of
interparticle interactions, specifically long-range dipole-dipole ones,
to the nanocomposite optical properties. Usually, the dipole-dipole
coupling in the system of nanoparticles is treated by numerical methods
\cite{Shalaev96,Myroshnychenko08}. This approach is very popular for
description of the systems of oscillating dipoles. Widely used approximations based on
homogenization (effective medium theories) have restrictions on the size of inclusions
\cite{Mackay11}, typical quantum dots do not fall in this range. In Ref.~\cite{Singh07}
mean-field theory was used in order to calculate the impact of dipole-dipole
interactions in the materials doped with five-level nanoparticles. It is
worth mentioning that the mean-field approaches are only suitable for the
medium comprising a set of dipoles with some ordering and for the nonzero mean field. In
this study, we obtain analytical expressions describing the effect of
the dipolar interactions on the nonlinear dielectric susceptibility of
quantum dot planar systems.

In these nanocomposites, incident laser light can excite the  electric dipole moment
in the nanoparticles. Upon increasing the
quantity of the dipolar excited nanoparticles, the dipole-dipole
interactions in the nanocomposite become of particular importance. To
calculate the contribution of the dipolar interactions to nonlinear
optical characteristics of the two-dimensional system of randomly
positioned cylindrical nanoparticles, the approach developed in
Ref.~\cite{Panov10} can be used. In this approximation, it is assumed
that the mean value $p$ of the induced electric dipole moment of the
particle is proportional to the electric field amplitude $E_{in}$ of the
incident (external) radiation,
\begin{equation}
p = \varepsilon_m \alpha v E_{in},
\label{p:Ein}
\end{equation}
where $\alpha$ is the dimensionless polarizability of the particle, $v$ is its volume, $\varepsilon_m$ is the dielectric function of ambient medium. Here $p$ denotes the absolute value of the particle dipole moment $\mathbf{p}$ which can be arbitrarily oriented in three dimensions. This continuous approximation can be justified for the quantum dots possessing a number of exciton levels. 
The experimentally measured values of the polarizabilities of the CdSe quantum dots in terahertz or static electric fields lie within the range from 0.4 (Ref.~\cite{WangF06}) to 3.6 (Ref.~\cite{Empedocles97}). The lifetime of the quantum dot excited state with dipole moment $\mathbf{p}$ is several orders greater than the period of an optical field oscillation. After the period much larger than the exciton rise time, typically ranging from tenths of picosecond \cite{Wang06,Wundke99} 
to tens of nanoseconds \cite{Thuy10}, a great number of the particles will be in the excited state and the system achieves a steady state. In the steady state, we suppose that the average quantity of the particles with the dipole moment does not vary and the
system reaches the thermodynamic limit. Thus, on a timescale of the optical field oscillation, the polarized particles can be regarded as static dipoles. By this means, the quantum dots with the dipole moment will induce on the test particle, being located at the coordinate origin, additional random electric field
\[
\mathbf{E}=\sum_l\frac{3(\mathbf{r}_l\cdot\mathbf{p}_l)\mathbf{r}_l-\mathbf{p}_lr_l^2}{\varepsilon_m r_l^5},\]
where $\mathbf{r}_l$ is the position vector of the $l$-th particle, $\mathbf{p}_l$ is its dipole moment. In polar coordinates, $\mathbf{r}_l$ has two components $(r_l,\phi_l)$. Hereinafter, we use the orientation of the test particle dipole moment as the selected direction. As a first approximation, we assume that the local field $\mathbf{E}$ or induced nonlinearities do not change polarizability $\alpha$ of the individual particles. This field is random since the quantum dots are polarized arbitrarily and placed haphazardly.
After obtaining the probability distribution function $W(E)$ for the projection $E$ of the field $\mathbf{E}$ onto the selected direction, the contribution of the dipole-dipole interactions to the nonlinear dielectric susceptibility of the nanocomposite can be calculated with the help of the statistical mechanics methods~\cite{Panov10}.

Let us consider the monolayer of cylindrical particles that lie randomly in a plane. These particles are assumed to have the electric dipole polarization (\ref{p:Ein}) after exposition to laser light and after the transition to the steady state. Due to the probabilistic nature of the excitation, the number of the dipoles with one direction is balanced by dipoles with opposite alignment. Thus, the ensemble lacks the ordering and the mean of the random field $\mathbf{E}$ is zero.
As shown in the Ref.~\cite{surfdip} for such a system, $W(E)$ becomes the Gaussian distribution at high nanoparticle surface concentrations $c=N s/S>0.6$, where $N$ is the number of the particles having been excited, $S$ is the area of the sample, $s$ is the area of the particle. Hence, we cannot apply directly the approach of Ref.~\cite{Panov10} exploiting the Gaussian distribution for the calculation of the third-order dielectric susceptibility as such high surface nanoparticle concentrations are hardly achievable in practice. In order to obtain $W(E)$ at $c<0.6$, making use of the negative cumulant expansion followed by the inverse Fourier transform was proposed~\cite{surfdip}:
\begin{equation}
W(E) =\frac{1}{2\pi}\int_{-\infty}^\infty\exp\left\lbrace-\frac{c}{s}\left[ \frac{1}{2}\lambda_2\rho^2-\frac{1}{4!}\lambda_4\rho^4 
+\frac{1}{6!}\lambda_{6}\rho^{6}+\ldots\right]-i\rho E  \right\rbrace \mathrm{d}\rho,
\label{wseriesh}
\end{equation}
where $\rho$ is the Fourier-transform variable and the cumulants $\lambda_n$ are defined as follows
\begin{equation}
\lambda_n=\frac1{p^n}\int \left\lbrace \frac{3(\mathbf{r}\cdot\mathbf{p})\mathbf{r}-\mathbf{p}r^2}{r^5}\right\rbrace ^n \tau(\mathbf{p})\mathrm{d} \mathbf{p}\, \mathrm{d}\mathbf{r},
\label{defmun}
\end{equation}
$\tau \left( \mathbf{p} \right)$ is the distribution function for dipole moment orientations. This approach is useful for the case of the identical functions $\tau \left( \mathbf{p} \right)$ for all the particles. Integration over radius $r$ in Eq.~(\ref{defmun}) begins from $2r_0$ which is the minimal possible distance between two centers of the cylindrical particles having a radius $r_0$. It should be emphasized that in the lack of the ordering in the system, cumulant expansion in Eq.~(\ref{wseriesh}) contains only even-numbered $\lambda_n$ \cite{surfdip}. Here, the absence of the ordering means that the net dipolar moment of the system is zero.

The free energy density  of the ensemble is found by the standard method:
\begin{equation}
\label{freeener}
 F=-k_B T  \frac{c}{v}\ln \int_{-\infty}^\infty \exp(-p E /k_B T)W(E) \mathrm{d}E,
\end{equation}
where $T$ is the temperature and $k_B$ is the Boltzmann constant, $v=sh$, $h$ is the width of the monolayer (height of the particle).
Projection $P$ of the sample macroscopic polarization onto the selected direction for monochromatic radiation and non-absorbing medium
can be obtained using the thermodynamic relation \cite{Shen84}
\begin{equation}
 P=-\frac{\partial \langle F\rangle_t}{\partial E_{in}},
 \label{therm_relP}
\end{equation} 
where $\langle \rangle_t$ denotes time averaging. It is worth recalling that in this model the dipole of the test particle can possess two opposite orientations with equal probabilities. To perform the time averaging, we check our results for the independence on the change of the sign of the exponential argument in Eq.~\ref{freeener} (Ref.~\cite{Panov10}). 

In the case of linear or circular light polarization, expanding $P$ in a series 
\begin{equation}
P=\chi_1 E_{in} + \chi_3 |E_{in}|^2 E_{in} + \ldots,
\label{def_chi3}
\end{equation}
it is possible to obtain cubic optical susceptibility of the system $\chi_3$ as well as higher order nonlinear susceptibilities.

Upon differentiating (\ref{therm_relP}) and expanding $P$ in a series, we get
\begin{eqnarray*}
\fl
P=\frac{c \alpha^2 v^2 \varepsilon_m^2}{ k_B T s h I_{0,0}^2}
\Biggl\lbrace \left[\frac{c}{s}\lambda_2 (k_B T)^2 I_{2,0}-I_{0,2}\right]I_{0,0}+I_{0,1}^2\Biggr\rbrace E_{in}+\\
\fl\quad
\frac{c \alpha^3 v^3 \varepsilon_m^3}{2 (k_B T)^2 s h I_{0,0}^3}\Biggl\{\left[ 3 \frac{c}{s} (k_B T)^2I_{1,2}\lambda_2-{I_{0,3}}\right]I_{0,0}^2 - \\
3\left[ \frac{c}{s}(k_B T)^2I_{0,2}\lambda_2-{I_{0,2}}\right]I_{0,1}I_{0,0} -2{I_{0,1}^3} \Biggr\}E_{in}^2+\\
\frac{c^2 \alpha^4 v^4 \varepsilon_m^4}{6 k_B T s^2 h I_{0,0}^4} 
\Biggl\{-
I_{0,0}^3\biggl[\left(\lambda_4+3\lambda_2^2\frac{c}{s}\right)(k_B T)^2 I_{4,0}-6I_{2,2}\lambda_2+\frac{s}{c(k_B T)^2}I_{0,4}\biggr]+ \\
4 I_{0,0}^2 I_{0,1} \left[\frac{sI_{0,2}}{c(k_B T)^2} - 3I_{2,0}\lambda_2\right]+
12 I_{0,0} I_{0,1}^2  \left[I_{2,0}\lambda_2 - \frac{sI_{0,2}}{c(k_B T)^2}\right]+\\
3 I_{0,0}^2 \frac{c}{s}\left[I_{2,0}\lambda_2 k_B T - \frac{sI_{0,2}}{c k_B T}\right]^2 + 6 I_{0,1}^4 
\Biggr\}E_{in}^3+\ldots,
\end{eqnarray*}
where
\begin{equation}
I_{k,l}=\int_{-\infty}^\infty \int_{-\infty}^\infty \exp(-i\rho E)\rho^k E^l \mathrm{d}\rho \mathrm{d} E
\label{Erhoint}
\end{equation}
To simplify these expressions, let us invoke the Fourier integral representation of the Dirac $\delta$-function \cite{Korn68}
\begin{equation}
\delta^{(n)}(E)=\frac{1}{2\pi}\int(i\rho)^n\exp(i\rho E) \mathrm{d}\rho
\end{equation}
and the definition of the Dirac $\delta$-function  $n$-th derivative \cite{Korn68}
\begin{equation}
\int f(E)\delta^{(n)}(E)\mathrm{d}E=(-1)^n f^{(n)}(0).
\end{equation}
Here $f$ is some function.
The use of the above two formulas reveals that integrals (\ref{Erhoint})
obtained after the expansion have nonzero values if $k=l$ and $k=0,2,4,\ldots$, namely $I_{0,0}=-2\pi$, $I_{2,2}=4\pi$.

In the absence of the ordering, the first non-vanishing term in expansion (\ref{def_chi3}) contains a third-order optical susceptibility $\chi_3(\omega;\omega,\omega,-\omega)$ (here $\omega$ is the cyclic frequency of the incident light).
After simplifications we derive formula for the self-induced Kerr nonlinear susceptibility:
\begin{equation}
\chi_3 = -\frac{ 2 c^2 \alpha^4 v^4 \varepsilon_m^4 \lambda_2}{k_B T s^2 h}.
\label{chi3-2d}
\end{equation}
 It should be noticed that the same relationship for $\chi_3$ can be obtained using the Gaussian distribution as $W(E)$.
The next nonzero term in expansion (\ref{def_chi3}) contains $\chi_7$:
\begin{equation}
\chi_7 = -\frac{ c^2 \alpha^8 v^8 \varepsilon_m^8 \lambda_4}{3(k_B T)^3 s^3 h}.
\label{chi7-2d}
\end{equation}
It should be noted that the above formulas are derived under condition $\varepsilon_m>0$. In the above calculations we did not limit the number of the terms in the negative cumulant expansion in (\ref{wseriesh}), so that the higher cumulants make contributions to higher-order susceptibilities.

Further, let us consider some special dipole configurations that can occur if the monolayer of the quantum dots is integrated into a planar waveguide or a surface polariton propagates along an interface with the nanoparticles. When all the dipoles have in-plane alignment and they are collinear with two equiprobable opposite directions,  the distribution function $\tau \left( \mathbf{p} \right)$ for dipole moment orientations is given by
\begin{equation}
\tau(\gamma)=\frac{\delta(\gamma)+\delta(\gamma-\pi)}{2},
\label{tauIsing}
\end{equation}
where the angle $\gamma$ is measured from the selected direction which is determined by the dipole orientation. Then, 
\begin{equation}
\lambda_2 = \frac{ 11 \pi }{16 \left( 2 r_0 \right)^4 \varepsilon_m^2}.
\label{mu2-2disingpar}
\end{equation}
For the case of the dipoles lying at the interface between two half-spaces with $\varepsilon_1$ and $\varepsilon_2$ \cite{Mohamed09},
\begin{equation}
\varepsilon_m = \frac{\varepsilon_1 + \varepsilon_2}{2}.
\end{equation}

For the in-plane oriented dipoles having arbitrary directions,
\begin{equation}
\tau(\gamma)=\frac 1{2\pi},\quad \lambda_2 = \frac{ 5\pi }{8 \left( 2 r_0 \right)^4 \varepsilon_m^2}.
\label{mu2-2dxy}
\end{equation}
Both of the above cases can occur when a TM-polarized wave propagates along the plane containing the nanoparticles.

Another situation arises when TE-polarized wave propagates along the interface.
In this case, the dipoles are expected to be oriented perpendicular to the plane. So we utilize function (\ref{tauIsing}) taking into account the fact that the polar axis is perpendicular to the plane. Then,
\begin{equation}
\lambda_2 = \frac{ \pi }{2 \left( 2 r_0 \right)^4 \varepsilon_m^2}.
\label{mu2-2disingperp}
\end{equation}
For the medium consisting of two half-spaces with $\varepsilon_1$ and $\varepsilon_2$ \cite{Mohamed09},
\begin{equation}
\varepsilon_m = \frac{2 \varepsilon_1 \varepsilon_2}{\varepsilon_1 + \varepsilon_2}.
\end{equation}
As one can see, all of these variants of $\lambda_2$ differ little. 

By way of illustration, in the case of the dipoles with in-plain collinear polarizations, substituting $v=sh$, $s=\pi r_0^2$, and (\ref{mu2-2disingpar}) in (\ref{chi3-2d}) we get
\begin{equation}
\chi_3 = -\frac{ 11 c^2 \alpha^4 h^3 \varepsilon_m^2}{128 k_B T}.
\label{chi3-2d-subs}
\end{equation} 
It is obvious that the obtained expression for the Kerr nonlinear susceptibility
does not depend explicitly on the particle radius. However, one should keep in
mind that the polarizability of the particles strongly depends on their size.
Provided $c=0.05$, $\alpha=0.4$, $h=5$~nm, $\varepsilon_m=2.3$, $T=300$~K using
Eq.~(\ref{chi3-2d-subs}) we arrive at $\chi_3 \approx -8\times 10^{-11}$~esu.
This value is of the same order as observed by experiment with nanosecond laser
pulses for cadmium chalcogenide nanocomposites
($-10^{-11}$--$-10^{-10}$~esu)\cite{Jing09,Mathew12}.

As well as for the bulk nanocomposite, the contribution of the dipole-dipole
interactions to the Kerr optical nonlinear susceptibility of the planar system
has a negative sign. This  is due to the disarranging effect of the random field
of the induced dipoles. It is worth mentioning that the dipolar optical
nonlinearity would become significant at the time intervals which exceed  the
exciton rise time many-fold. Comparison with the results of Ref.~\cite{Panov10}
shows that for the same $r$ and $h$ the contribution of the dipole-dipole
interactions to the Kerr susceptibility of the two-dimensional nanocomposite is
several times less than for the bulk nanocomposites. This is especially important for the
nanoparticles which acquire the large values of the dipole moment, for instance,
the II-VI semiconductor quantum dots \cite{Colvin92}. Often, under femtosecond
laser pulses, these quantum dots exhibit the positive real part of the
third-order susceptibility \cite{Gan08,Zhu11}, but at nanosecond pulse durations
they show negative nonlinear refractive index \cite{Jing09,Mathew12}. This
phenomenon may be attributed to the effect of the dipole-dipole interactions.

If the dipolar optical nonlinearity is undesirable, the organization of the
nanocomposites, which contain the nanoparticles with the great magnitudes of the
induced dipole moment, as planar structures is justified. However, one would 
take into account that other types of optical nonlinearity also may decrease
with reducing the number of dimensions.

As one can see from Eq.~(\ref{chi3-2d}), $\chi_3$ is proportional to $c^2$. It
is the dependence on concentration that is the key feature of the contribution
of the interparticle dipole-dipole interactions to the optical nonlinearity of
the composite. This is reasonable since the number of interparticle couplings
varies as the square of the quantity of the particles in the sample. As one
could expect, other types of optical nonlinearities would linearly depend on the
nanoparticle concentration.

It should be underlined that the proposed here approach to the calculation of
the dipole-dipole interaction contribution to the nonlinear susceptibility of
nanocomposite can be applied not only to the two-dimensional geometry but also
to other configurations. The method, based on the cumulant expansion of the
characteristic function, can be also utilized for other types of interparticle
interactions, such as the quadrupolar ones, since there is no need to determine
probability distribution function $W(E)$ explicitly.

In conclusion, the analytical expressions for Kerr optical nonlinearity $\chi_3$
of the two-dimensional system of the randomly located quantum dots have been
derived. It has been shown that for the different types of the light
polarization, $\chi_3$ changes insignificantly. The dipole-induced optical
nonlinearity of the two-dimensional sample is several-fold smaller than for the
bulk one so that this can be used in practice.

\clearpage
\raggedright
\bibliography{inter_part}
\end{document}